\documentclass{ws-procs11x85}
\usepackage{ws-procs-thm}           

\usepackage{graphicx}
\usepackage{multirow}
\usepackage{booktabs}
\usepackage{algorithmic}
\usepackage{mathtools}
\usepackage{microtype}

\begin{document}

\title{Network Representation of Large-Scale Heterogeneous RNA Sequences with Integration of Diverse Multi-omics, Interactions, and Annotations Data \\
}

\author{Nhat Tran and Jean Gao}
\address{Department of Computer Science \& Engineering, The University of Texas at Arlington,\\
Arlington, TX 76010, USA\\
E-mail: gao@uta.edu\\
}



\begin{abstract}
Long non-coding RNA (lncRNA), microRNA, and messenger RNA enable key regulations of various biological processes through a variety of diverse interaction mechanisms. Identifying the interactions and cross-talk between these heterogeneous RNA classes is essential in order to uncover the functional role of individual RNA transcripts, especially for unannotated and sparsely discovered RNA sequences with no known interactions. 
Recently, sequence-based deep learning and network embedding methods are gaining traction as high-performing and flexible approaches that can either predict RNA-RNA interactions from sequence or infer missing interactions from patterns that may exist in the network topology. However, most of the current methods have several limitations, e.g., the inability to perform inductive predictions, to distinguish the directionality of interactions, or to integrate various sequence, interaction, expression, and genomic annotation datasets. 
We proposed a novel deep learning framework, rna2rna, which learns from RNA sequences to produce a low-dimensional embedding that preserves proximities in both the interaction topology and the functional affinity topology. In this proposed embedding space, the two-part \textquotedblleft source and target contexts\textquotedblright capture the receptive fields of each RNA transcript to encapsulate heterogeneous cross-talk interactions between lncRNAs and microRNAs. The proximity between RNAs in this embedding space also uncovers the second-order relationships that allow for accurate inference of novel directed interactions or functional similarities between any two RNA sequences. 
In a prospective evaluation, our method exhibits superior performance compared to state-of-art approaches at predicting missing interactions from several RNA-RNA interaction databases. Additional results suggest that our proposed framework can capture a manifold for heterogeneous RNA sequences to discover novel functional annotations.
\end{abstract}
\copyrightinfo{\copyright\ 2019 The Authors. Open Access chapter published by World Scientific Publishing Company and distributed under the terms of the Creative Commons Attribution Non-Commercial (CC BY-NC) 4.0 License.}

\section{Introduction}
Regulatory long non-coding RNAs (lncRNAs) and microRNAs (miRNAs) that influences gene expression post-transcriptionally by interacting to target messenger RNAs (mRNA) form a complex network of transcriptomic interactions. These heterogeneous families of non-coding RNAs are associated with nearly all cellular processes, and recently, lncRNAs are gaining considerable attention as the largest and most diverse class of non-coding RNA, encompassing nearly 30,000 discovered transcripts in human. Among many of their known functional interaction mechanisms, lncRNAs are known to act as miRNA decoys, derepress gene expression by competing with miRNAs for shared mRNA targets, or directly regulate gene expression \cite{yoon2014functional}. 
Determining the biological functions of the individual lncRNAs remains a challenge as most of these RNA transcripts are currently unannotated, and their known interactions are sparse. Recent advances in RNA sequencing (RNA-Seq), deep sequencing (CLIP-seq, LIGR-Seq), and computational methods allow for an unprecedented analysis of such transcripts and have enabled researchers to generate large-scale interaction and functional annotation databases. However, the interaction networks generated from such data are often scant and incomplete in the number of lncRNAs covered. Furthermore, although a large number of lncRNAs have been identified, only a few hundreds have had functional and molecular mechanisms determined to date, as observed in lncRNAdb \cite{amaral2010lncrnadb}. In other avenues, a growing number of lncRNAs are being assigned biological functions based on their cell-specific expressions \cite{gloss2016specificity} or disease associations \cite{chen2015constructing}. Still, a vast majority of other lncRNAs only has basic genomic information such as locus biotype and transcript sequence assigned.
These transcripts might support important biological cell functions and could potentially serve as targets for genomic, diagnostic, or therapeutic studies. Thus, in the effort to functionally characterize these hypothetical lncRNAs, the essential tasks are integrating the various -omics and annotation attributes between the known RNAs, while inferring the functions and interactions to the novel RNAs directly from sequence. In order to address these challenges, this paper proposes a \textquotedblleft network embedding\textquotedblright method, which aims to provide a network representation of existing knowledge in the interaction topology and attributes of transcriptome-wide RNAs, all while predicting novel functional interactions. 

Many graph-theoretic methods have been applied to biological networks with the intuition that RNAs close together in the interaction topology are more likely to be involved in many of the same functions \cite{chua2006exploiting}. Due to the extreme sparsity of the known interaction network among lncRNAs, it is pertinent to unravel the functional association between lncRNAs by considering its gene/transcript genomic annotations, tissue-specific expressions, gene-disease associations, and sequence similarity \cite{chen2015constructing}. In recent studies, Kishan \textit{et al.} \cite{kishan2019gne} uncovered the second-order proximity relationship between interacting genes by integrating the gene regulatory network and gene expression as side information. On the other hand, several structure-free sequence-based methods \cite{asgari2015continuous, quang2016danq, pan2018learning} have also been proposed for the prediction of protein binding sites, family classification, structure prediction, or interaction prediction from sequence. Motivated by these works, we propose an integrative method to extract a functional representation from RNA sequences in order to accurately predict new interactions and functional annotations, while simultaneously integrate various existing biological multi-omics and annotation databases. 
\section{Methods}
\subsection{Defining the Heterogeneous lncRNA-miRNA-mRNA Network}
We formally define the heterogeneous network of lncRNA, miRNA, and mRNA interactions and functional similarity as two networks of directed and undirected edges. We denote the two networks $G_1(V, E^d)$ and $G_2(V, E^u)$ having the same set of nodes $V$ and two sets of edges $E^d$ and $E^u$. The set of nodes $V=\{v_1,\ldots,v_n\}$ can also be expressed as $V=\{L,M,N\}$ s.t. $|L|+|M|+|N| = n$, where $L, M, N$ are the sets containing the lncRNA, microRNA, and mRNA heterogeneous nodes, respectively.
The set of directed edges $E^d=\{e^{d}_{ij}\}^{n}_{i,j=1}$ represent directed regulatory interactions that each specify a source and a target.
The undirected edges $E^u=\{e^{u}_{ij}\}^{n}_{i,j=1} \mathrel{s.t.} e^{u}_{ji}=e^{u}_{ji}$ represent the functional affinities among the heterogeneous RNA nodes. Each edge $e_{ij}$ is associated with a weight $\mathrel{s.t.} 0 \leq e_{ij} \leq 1$, indicating the strength of the connection between RNA $i$ and RNA $j$. If $e_{ij} > 0$, we consider the edge a positive interaction/affinity, and if $e_{ij} = 0$, we consider the edge a negative (non) interaction/affinity. In this paper, we mainly consider the edge weights $e_{ij}$ to be binary.

\subsection{Materials and Data}
The directed edges represent the directed regulatory interactions between lncRNAs, miRNAs, and mRNAs. We consider the directionality in the regulatory interactions by encoding directed edges, i.e., encoding miRNA-lncRNA interactions (e.g., miRNA inducing lncRNA decay) to be separate from lncRNA-miRNA interactions (e.g., lncRNAs acting as miRNA decoys). In this study, the various types of interaction collected from various experimentally-verified databases in the lncRNA-miRNA-mRNA interactome considered are:
\begin{itemlist}
	\item \textbf{microRNA-mRNA} post-transcriptional interactions from miRTarBase \cite{chou2017mirtarbase} v7.0 containing a total of 442,067 interactions matched between 1,630 microRNAs and 14,666 mRNAs.
	\item \textbf{microRNA-lncRNA} interactions from DIANA-lncBase Experimental v2 \cite{paraskevopoulou2015diana} containing a total of 53,926 matched interactions between 631 miRNAs and 2530 lncRNAs.
	\item \textbf{ncRNA-RNA} regulatory interactions from NPInter v3.0 \cite{hao2016npinter}, where we filtered only lncRNA-miRNA, lncRNA-mRNA, and miRNA-lncRNA interactions, which resulted in 208,841 interactions between 669 matching lncRNAs and 7,369 mRNAs.
	\item \textbf{lncRNA-mRNA} post-transcriptional regulatory interactions from lncRNA2Target v2.0\cite{cheng2018lncrna2target} which contained a total of 65,655 interactions between 1037 lncRNAs and 28,866 mRNAs.
	\item \textbf{mRNA-mRNA} gene regulatory interactions obtained from the BioGRID v3.5 database \cite{chatr2017biogrid}, which included more than 347,246 matched interactions among 19,607 mRNAs. 
\end{itemlist}
These heterogeneous interactions are combined into an integrated network, and the associated set of edges is $E^d$, where the binary edge weight $e^d_{ij} \in \{0,1\}$  indicates whether a regulatory interaction from RNA node $v_i$ to RNA node $v_j$ has been observed in the literature. 

We also outline the various multi-omics, sequence, and annotation databases utilized to provide functional attributes to individual RNAs. In order to maximize the number of transcripts and genes matched between different annotation systems, the RNA transcripts are indexed by standard gene symbols.
\begin{itemlist}
	\item \textbf{Tissue-specific median expressions} obtained from the GTEx Portal \cite{lonsdale2013genotype}, which provides tissue-specific RNA-Seq expressions for 15,598 non-diseased samples of various human tissue types. From GTEx Analysis v7, median TPM values by tissue were calculated for 18653 mRNA genes, 1181 miRNA and 12706 lncRNA transcripts over 44 tissue types. 
	\item \textbf{Genomic annotations} obtained from the GENCODE Release 29 \cite{harrow2012gencode} which contains the transcript biotype annotation and genomic location for lncRNAs, miRNAs, and mRNAs. 
	\item \textbf{Transcript sequences} comprised of genome-wide human lncRNA and mRNA primary sequences from GENCODE Release 29 \cite{harrow2012gencode} and miRNA hairpin sequences from miRBase \cite{griffiths2006mirbase}.
	\item \textbf{Functional annotations} for mRNA were obtained from the HUGO Gene Names database \cite{eyre2006hugo}. In addition, GO terms for 162 matched lncRNAs were obtained from RNAcentral \cite{rnacentral2016rnacentral} which aggregated data from NONCODE \cite{bu2011noncode} and lncipedia \cite{volders2012lncipedia}. For miRNAs, family classifications obtained from the TargetScan Release 7.2 (March 2018) \cite{agarwal2015predicting}, RNA structure family annotation from Rfam 13.0 \cite{kalvari2017rfam}, and GO terms for 487 matched miRNAs from RNAcentral were included. 
	\item \textbf{Disease associations} from DisGeNet \cite{pinero2016disgenet} covered 7577 mRNA genes. The HMDD miRNA-disease\cite{huang2018hmdd} database includes 553 miRNAs, and LncRNADisease \cite{chen2012lncrnadisease} includes 150 lncRNAs. 	
\end{itemlist}
Each of the RNA functional attribute data types, excluding transcript sequences, is encoded to a data matrix, $A^k \in \Re^{|V| \times m}$ for the annotation type $k$, which can either be real-valued or categorical. If an attribute $k$ is categorically-valued, $A^k$ is an indicator matrix where entries are binary-valued, and $m$ is the number of all unique values. Considering that functional annotations for the non-coding RNAs are sparse, each node can have up to $K$ attribute types associated with it.  

\subsection{Undirected RNA-RNA Functional Affinity Edges}
We aim to capture the functional similarity between two RNA nodes of the same class by calculating as a similarity measure, suggesting a resemblance in RNA function or structure. For any categorical text-valued annotation (e.g., disease association, transcript biotype, RNA structure family, or GO terms) that two RNA nodes $v_i$ and $v_j$ both have been annotated, the attributes in this annotation are first transformed to binary feature representation. For a categorical annotation denoted as $k$, the binary 1-D feature vector associated with RNA node $v_i$ is denoted as $a^k_i$. Using the S{\o}rensen-Dice coefficient score \cite{dice1945measures}, a similarity score between two binary vectors for node $v_i$ and node $v_j$ for annotation $k$ can be obtained by $s^k_{ij}= {2(a^k_i \cdot a^k_j)}/(2(a^k_i \cdot a^k_j)+ |a^k_i|_1 + |a^k_j|_1)$. 
This similarity measure ranges $[0, 1]$ and gives higher weight to the common attributes present in both RNAs than by the attributes present in only one RNA. Since most RNAs have null annotations, the Dice coefficient score is only computed between pairs of RNA nodes that have both been annotated, while other pairs remain a null value. 
For the real-valued attribute data, i.e., tissue-specific expressions, we instead calculate the absolute correlation value between the two expression vectors $a^k_i$ and $a^k_j$ to obtain $s_{ij}^k$. 

To obtain an aggregate affinity score between a pair of RNA nodes across all $K$ similarity values, we utilize a modified version of the Gower's similarity score \cite{gower1971general}. For each RNA-RNA pair, Gower's similarity aggregates the similarity scores across all the annotation types and performs a weighted average. Typically, a similarity score for a pair of RNA nodes that do not have any associated annotation would be considered a $0$, but in this study, we remove these null pairwise similarity from consideration. Thus, Gower's similarity will only aggregate the pairwise similarity scores from annotations that exist in both nodes to compute the average. 

After computing for all pairs of RNA nodes, the resulting pairwise affinity matrix is $\mathbf{S}$ where entries $S_{ij}=\sum_k^{K_{ij}}{ s^k_{ij}}/{|K_{ij}|} $ with $K_{ij}$ being the set of attribute types present in both RNA nodes $v_i$ and $v_j$. The entries $S_{ij}$ will then be selected as edge weights $e^u_{ij}$ that represent the functional similarity edges between nodes. Since our model currently only considers unweighted binary edges, we selected undirected edges to have an affinity score close to $1.0$, or higher than a chosen hard-threshold, to be considered as a positive edge. In our experiments, the hard-threshold was arbitrarily chosen where the number of positive affinity edges covers no more than $0.1\%$ sparsity of the entire affinity matrix. We also utilize the set of undirected affinity edges with $0.0$ weight, indicating a negative edge that suggests functional dissimilarity.



\section{Network Embedding with Source-Target Contexts}
A network embedding is mapping from each RNA node to a low-dimensional representation, denoted as a mapping function $f: v_i \rightarrow y_i \in \Re^d, \forall v_i \in V$, where $d$ is the embedding dimensionality $\mathrel{s.t.} d \ll n$. The embedding $y_i$ associated with each node $v_i$ is learned such that nodes preserve some meaningful proximities to each other according to the given topology in both networks $G_d$ and $G_u$. 
We propose the embedding to have two components: source context and target context. That is, each embedding vector $y_i = [s_i, t_i]^\top$ is represented as a concatenation of the source context, $s_i \in \Re^{d/2}$, and target context, $t_i \in \Re^{d/2}$. This embedding representation can simultaneously capture directed and undirected edges by our proposed first-order directed proximity and second-order undirected proximity.
The first-order directed proximity represents a binding affinity from node $i$'s source context to node $j$'s target context, defined as 
\begin{equation}
	d_1(v_i, v_j) = \sqrt{(s_i - t_j)^2}	
	\label{eq_d1}
\end{equation}
Note, $d_1(v_i, v_j)$ can take on a different value than $d_1(v_j, v_i)$.
The second-order undirected proximity represents the functional similarity between node $i$ and node $j$, defined as 
\begin{equation}
	d_2(v_i, v_j) = \sqrt{(y_i-y_j)^2}
\end{equation}
The value of these proximities is a Euclidean distance, where the embeddings are selected so that if two nodes have a positive (directed or undirected) edge, its respective embeddings should be more similar, i.e., having a smaller distance. Otherwise, if two nodes have a negative or non-interactions, their embeddings should be more dissimilar, incurring a greater distance.

\begin{figure*}[t]
\centerline{\includegraphics[width=\textwidth]{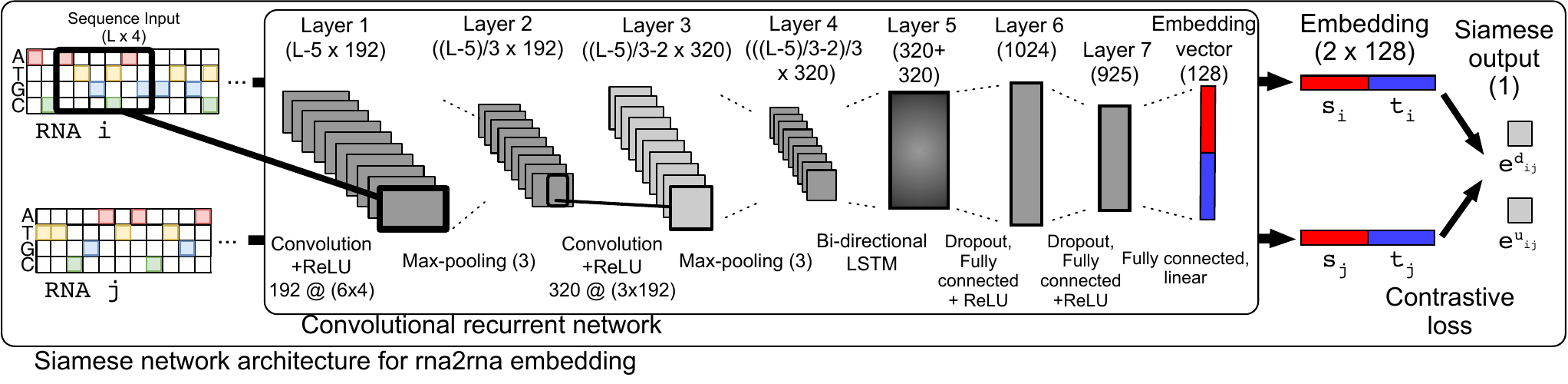}}
\caption{The rna2na network embedding method utilizing Siamese architecture.}
\label{architecture}
\end{figure*}

\subsection{Representation Learning for RNA Sequences to Reconstruct the Interactions and Functional Topology}

Aside from the interaction topology data, each RNA $v_i$ also has an associated transcript sequence in one-hot vector representation denoted by $x_i \in \{1,2,3,4\}^{l_i}$, where $l_i$ is the variable length of an RNA sequence. We propose the network embedding function $f : x_i \rightarrow y_i \in \Re^d$ to be a deep Siamese neural network that maps an RNA sequence input $x_i$ to an embedding $y_i$ of a fixed dimension $d=128$, outlined in Fig. \ref{architecture}. 
Initially proposed for signature verification \cite{bromley1994signature}, the Siamese network is an architecture where a pair of objects can each be encoded such that its resulting embeddings can determine if the two objects are similar or dissimilar. More specifically, the network learns to output embeddings for a pair of RNA sequences, guided by edge weight $e^u_{ij}$ as the label indicating whether the pair is functionally similar or dissimilar. Additionally, for an interacting pair of RNAs, the directed edge $e^d_{ij}$ would indicate whether RNA $i$ interacts with RNA $j$. For the output embeddings to preserve proximities across all edges in both $G_d$ and $G_u$ network topologies, we minimize the binary cross-entropy losses \cite{koch2015siamese}:
\begin{equation}
	L_1(X, E^d, f) = \sum_{e^d_{ij} \in E^d} e^d_{ij} \log(d_1(f(x_i), f(x_j))) +(1 - e^d_{ij}) \log(1-d_1(f(x_i),f(x_j)))
\end{equation}
\begin{equation}	
	L_2(X, E^u, f) = \sum_{e^u_{ij} \in E^u} e^u_{ij} \log (d_2(f(x_i), f(x_j))) + (1-e^u_{ij}) \log(1-d_2(f(x_i),f(x_j)))
\end{equation}
\label{batch_sampling}
The network weights in $f(x)$ are trained with Stochastic Gradient Descent using the standard back-propagation algorithm. At each iteration, a batch of RNA nodes is sampled along with its associated sets of positive and negative, directed and undirected edges. This batch optimization strategy addresses the memory requirement to encode a large number of variable-length RNA sequences while efficiently sample the associated interaction edges. Additionally, it also circumvents the issue of size imbalance between the interaction sets of well-studied RNA classes and newly-emerged RNA classes. We implement a biased random node sampling strategy where we randomly select a set of nodes and fit the network on the set of edges induced by this sub-graph. The probability of selecting a node $i$ is a function of its degree $r_i$, as a square-root sampling compression function \cite{riad2018sampling} $P(v_i) = \sqrt{r_i} / \sum_{v_j\in V} \sqrt{r_j}$, which aims to retain the power-law degree distribution while keeping the frequency ranking of each node. A batch of nodes $S$ is sampled without replacement from this distribution, and each node has a set of positive edges. To obtain the negative directed edges, we then sample a number of nodes in $S$ by adopting the negative sampling approach proposed in \cite{mikolov2013distributed}, where the ratio of negative edges to positive edges incident to each node is between $2.0$ and $5.0$.
Given $S$ and the set of directed $E^d_S$ and undirected $E^u_S$ edges containing both positive and negative interactions incident to $S$, we minimize the loss function
$L(S, E^d_S, E^u_S, f) =  L_1(S, E^d_S, f) + \lambda L_2(S, E^u_S, f)$
where $\lambda$ is a coefficient parameter.

After training convergence, given any RNA full sequence inputs $x_i$ and $x_j$, the learned model yields the embeddings $y_i$ and $y_j$ which is used to predict whether a relationship exists between them. We use the proximity $d_1(v_i,v_j)$ or $d_2(v_i,v_j)$ to either predict the existence of interaction or functional similarity, depending on the query. Finally, a Gaussian kernel function $P(v_i,v_j) = \exp(-\gamma * d(v_i, v_j)^2)$ is applied to yield a predicted affinity score. 

\section{Results}
\subsection{Large-scale Integration of RNA Sequences, Interactions, Expressions, and Annotations}
We integrated various experimentally verified interaction databases to build a large-scale lncRNA-miRNA-mRNA interaction network. Additionally, various functional annotations, expressions, sequences, and disease associations were integrated to enable the extraction of the undirected attribute affinity edges. In total, there are 12725 lncRNAs, 1870 microRNAs, and 20284 mRNAs considered in this study, comprised of comprehensive integration of the various databases outlined in Table \ref{interaction_databases}. 
\begin{table*}[t]
\tbl{Overview of the data selection, harmonization, and integration for training and validation interaction databases in prospective evaluation.}
{\begin{tabular}{@{}l|rrrr@{}}
\toprule
	\multicolumn{1}{l|}{\bfseries Interaction database} &
	\multicolumn{4}{c}{\bfseries Training Sets} \\ 
\cline{2-5}
\bfseries & 
\bfseries Version & 
\bfseries \# interactions & 
\bfseries \# source nodes & 
\bfseries \# target nodes \\
\hline
miRTarBase & 6.0 & 377,318 & 1,618 miRNAs & 14,666 mRNAs \\
DIANA-lncBase & v2 & 53,926 & 631 miRNAs & 2530 lncRNAs \\
NPInter & v2.0 & 85,335 & 12 lncRNAs & 5023 mRNAs \\
lncRNA2Target &  v1.0 & 1308 & 79 lncRNAs & 471 mRNAs \\
BioGRID & v3.4 & 313,724 & 13,318 mRNAs & 19,429 mRNAs \\   
\hline
\end{tabular}}

\quad \quad \\
\begin{tabular}{@{}l|rrrr@{}}
\toprule
	\multicolumn{1}{l|}{\bfseries Interaction database} &
    \multicolumn{4}{c}{\bfseries Validation Sets} \\ 
\cline{2-5}
\bfseries & 
\bfseries Version &
\bfseries \# interactions &
\bfseries \# novel sources & 
\bfseries \# novel targets \\
\hline
miRTarBase & 7.0 & 64,749 & 12 miRNAs & 702 mRNAs \\
DIANA-lncBase & Predicted & 337,031 & 0 miRNAs & 0 lncRNAs \\
NPInter & v3.0 & 123,054 & 499 lncRNAs & 2346 mRNAs \\
lncRNA2Target & v2.0 & 65,624 & 1037 lncRNAs & 10,825 mRNAs \\
BioGRID & v3.5 & 33,522 & 178 mRNAs & 178 mRNAs \\   
\hline
\end{tabular}
\label{interaction_databases}
\end{table*}
To accomplish the primary task of predicting novel interactions not seen at training time, we propose an experimental setup using prospective evaluation. All models were trained exclusively using a prior version of each interaction database. Then, we validate the link prediction model using the set of new interactions from the latest database version update. This type of evaluation, rarely done in the literature, is extremely important as it allows us to mimic a realistic scenario where the task is to discover novel RNA-RNA interactions based on the existing knowledge. 
The training sets are comprised of the interaction network of database versions released before 2015, while the validation sets are comprised of updates from the latest released database version. After integration of the training databases, self-interactions and redundant interaction edges are removed, and only interactions between RNAs with an associated transcript sequence is considered. In the validation set, we evaluated only the interactions that do not overlap with the training set.

After the integration of various multi-omics and annotation attributes, RNA-RNA pairwise functional affinities were computed, and undirected affinity edges were added to the undirected interactions training set. Given the affinities $\mathbf{A}$ for all lncRNA, miRNA, and mRNA pairs and filtering second-order undirected affinities at a $0.8$ threshold, 65864, 405, and 362362 undirected edges were added, respectively. With the negative sampling ratio set at $5.0$ per positive edge, a total of 329320, 2025, and 724724 negative edges among the lncRNAs, miRNAs, and mRNAs were added to the undirected edges training set, respectively.

\subsection{Novel Link Predictions}
Our experiments included comparative analysis with existing state-of-the-art network embedding methods across various evaluation tasks. The methods considered are node2vec \cite{grover2016node2vec}, LINE \cite{tang2015line}, HOPE \cite{ou2016asymmetric}, and SDNE \cite{wang2016structural}. In the following experiments, we assessed each method by learning a 128-dimensions embedding representation from the training network. All other free parameters were set according to the default values in the proposed methods.
\begin{figure*}[t]
\begin{tabular}{c}
	\centerline{
  \includegraphics[width=33mm]{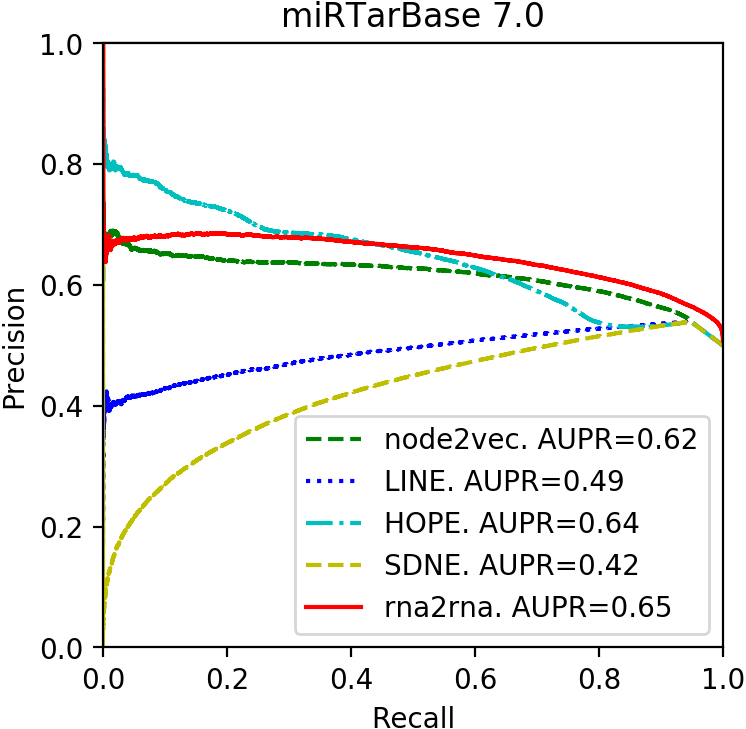}
  \includegraphics[width=33mm]{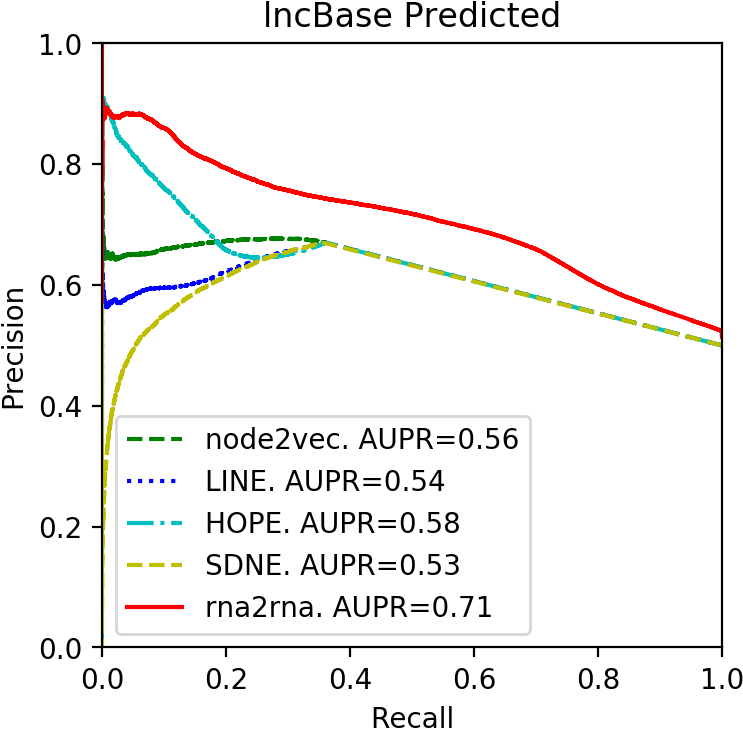}
    \includegraphics[width=33mm]{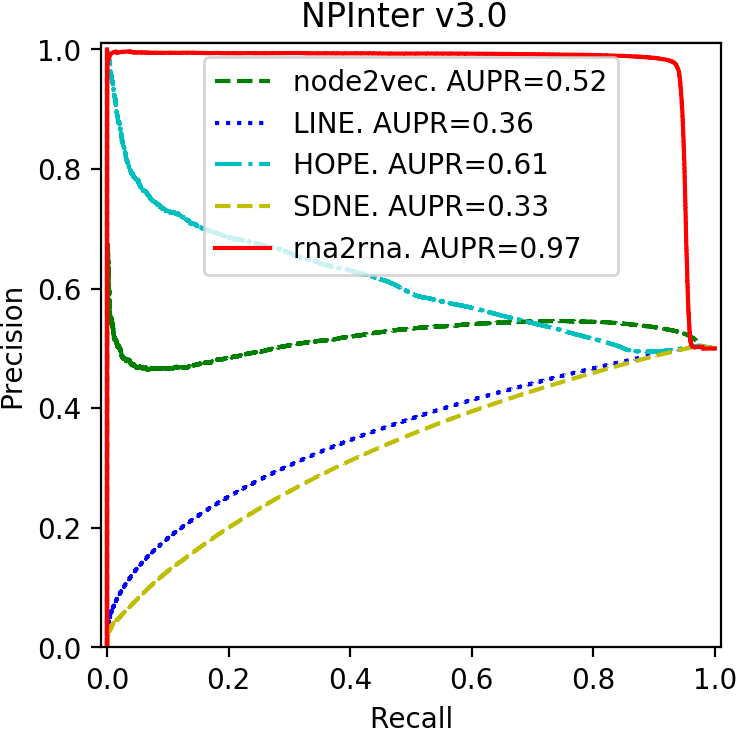}
 \includegraphics[width=33mm]{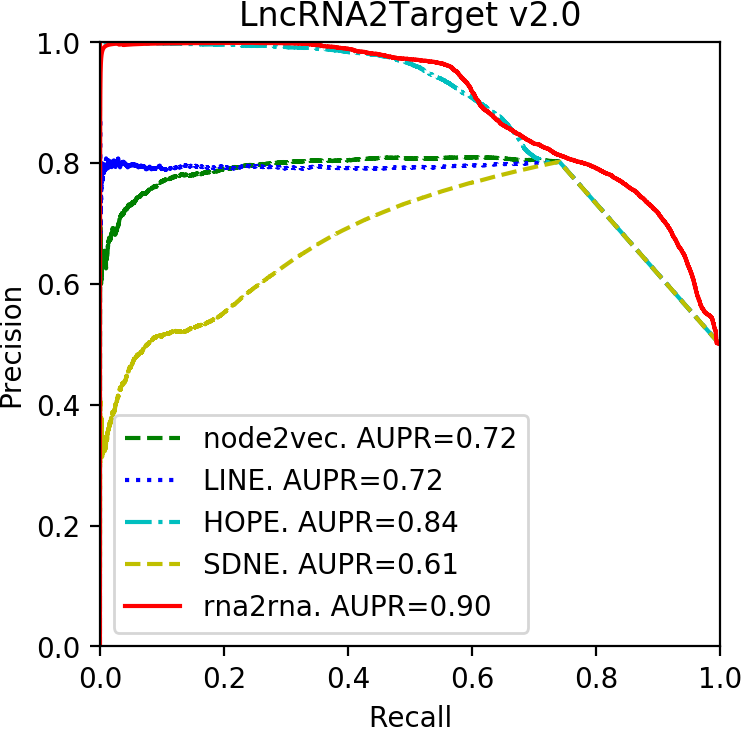}
 \includegraphics[width=33mm]{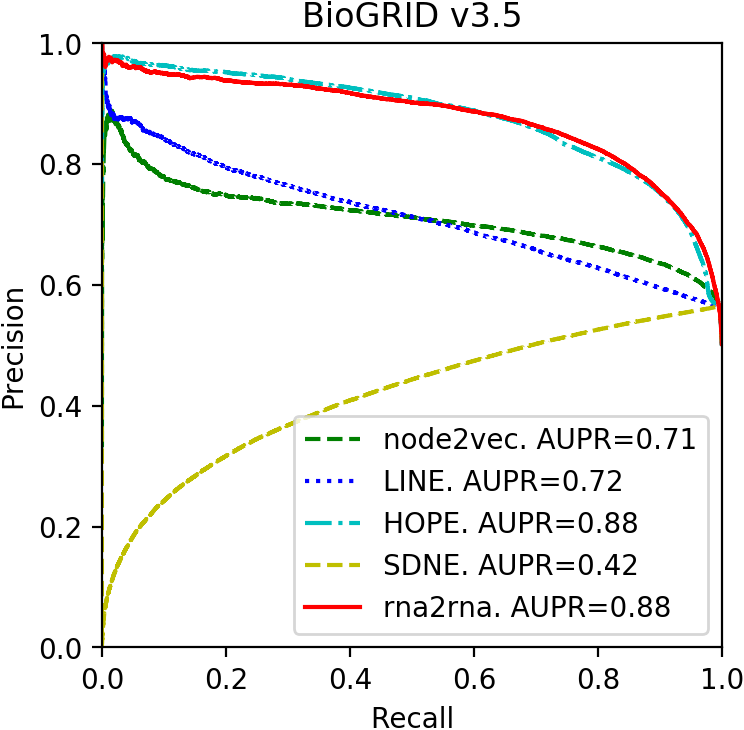}} \\
 
\end{tabular}
\caption{Precision-recall curves for link prediction comparison analysis. 
}
\label{pr_link_prediction}
\label{fig:node1_10}
\end{figure*}
We composed the training set for the link prediction task by a union of all ground-truth interactions set from the miRTarBase 6.0, lncBase v2, NPInter v2.0, lncRNA2Target v1.0, and BioGRID v3.4 databases. After the models have been trained, its estimated interaction adjacency matrix is computed and evaluated on the novel interactions from miRTarBase v7.0, lncBase predicted, NPInter v3.0, lncRNA2Target v2.0, and BioGRID v3.5 database separately.
For a test to differentiate between positive interactions and random noise interactions, we also uniformly sample several interactions from the set of all possible pairwise interactions to consider as negative interaction using the random node sampling distribution specified in Riad \textit{et al} \cite{riad2018sampling}.
This set is denoted as $E^n$, and sampling size of negative interactions is such that the ratio of negative to positive interactions is $1.0$.
In the evaluation, the sets of ground truth validation edges $E^d$ and random noise $E^n$ edges are used to calculate the precision and recall rates. All methods were evaluated on the same set of positive and sampled negative interactions. The area under the precision-recall (AUPR) curves shown in Fig. \ref{pr_link_prediction} highlights the comparison analysis across five different interaction databases. Note that rna2rna can perform inductive link prediction to novel RNA sequences since the validation set contains many novel lncRNAs, miRNAs, and mRNAs not seen at training time. While other transductive methods were evaluated only on the interactions incident to the RNA nodes seen at training time, rna2rna were evaluated on a greater number of interactions, yet still achieved the top performance at the default parameters $\gamma=2.0$ and $\lambda=1.0$.

\subsection{Inferring Functional Similarity From Embeddings}
In comparison analysis, we first obtained the embeddings from each of the methods and performed K-Means clustering only on the nodes that have an associated functional annotation. The number of clusters in K-Means is the same as the total number of unique labels in a particular annotation. The clustering result of different methods is compared over the RNA family and transcript biotype annotations, shown in Table \ref{clustering_family} and Table \ref{clustering_type}. The result shows that although there is a greater number of RNA nodes to assign to clusters, rna2rna embeddings can achieve the highest NMI score over the RNA functional family and biotype annotations.


\begin{table}[t]
\centering
\setlength\tabcolsep{2pt} 
\parbox{.45\linewidth}{
\tbl{Clustering analysis over 2343 true RNA functional family annotations.}
{\begin{tabular}{lcccr}
\toprule
Method & \scriptsize Homogeneity & \scriptsize Completeness & NMI & \# nodes \\
\colrule
node2vec & 0.641 & 0.602 & 0.621 & 11735 \\
LINE & \textbf{0.689} & 0.614 & 0.650 & 11735 \\
HOPE & 0.525 & 0.571 & 0.570 & 11735 \\
SDNE & 0.613 & 0.588 & 0.600 & 11735 \\
rna2rna* & 0.508 & 0.530 & 0.519 & 14312 \\
rna2rna & 0.685 & \textbf{0.620} & \textbf{0.651} & 14312 \\
\botrule
\end{tabular}}\label{clustering_family}
}
\quad \quad \quad
\parbox{.45\linewidth}{
\tbl{Clustering analysis over 24 true RNA transcript biotype annotations.}
{\begin{tabular}{lcccr}
\toprule
Method & \scriptsize Homogeneity & \scriptsize Completeness & NMI & \# nodes \\
\colrule
node2vec & 0.147 & 0.089 & 0.111 & 23940 \\
LINE & 0.268 & 0.158 & 0.199 & 23940 \\
HOPE & 0.109 & 0.111 & 0.110 & 23940 \\
SDNE & 0.079 & 0.076 & 0.078 & 23940 \\
rna2rna* & 0.178 & 0.138 & 0.155 & 32530 \\
rna2rna & \textbf{0.355} & \textbf{0.235} & \textbf{0.283} & 32530 \\
\botrule
\end{tabular}}\label{clustering_type}
}
{\scriptsize rna2rna* denotes the model trained on the directed interactions data alone, without the undirected functional affinity information.}
\end{table}

Since rna2rna embeddings can preserve functional similarities, an essential next step is to assign putative biological functions to novel lncRNAs. To do this, we perform gene set enrichment analysis on K-mean clusters of the RNA node embeddings, select the clusters with the highest enriched functional term, then associate the lncRNAs belonging in this cluster with this term. In this experiment, K-means ($k=2000$) is performed over 32,741 RNA embeddings trained from both training set and validation set. We then performed enrichment analysis on these 2000 clusters using Enrichr \cite{kuleshov2016enrichr} over the KEGG Human 2019 \cite{kanehisa2000kegg} terms, which includes both functional and disease pathways. Among the 2000 clusters, 559 have an adjusted P-value of less than $0.01$, and 139 have less than $0.001$. Interestingly, the highest-scoring gene sets usually contain some lncRNAs not previously associated with any other functional term, as shown in Table \ref{cluster_gsea}. It warrants additional experimental studies to verify the functional associations of these lncRNAs to the annotated genes within the same cluster.




\subsection{Discussions and Conclusion}
With the framework we have developed, heterogeneous functional attributes and interaction data are integrated to enable the characterization of RNA sequences using a network embedding representation.
While the method of integrating various functional attributes is simple, its purpose is to characterize functional affinities for an extensive number of RNAs, even among sparsely annotated ones. Whereas very few lncRNAs have been fully annotated, especially in functional annotation or disease association, most have already been annotated with tissue-specific expressions, transcript biotype, and sequence. Thus, more training data was utilized so the model can recognize a greater number of RNA sequences to achieve better generalization.

\begin{table*}[t]
\setlength\tabcolsep{3pt} 
\tbl{Gene set enrichment analysis over 2000 K-mean clusters. Each row indicates the highest enriched KEGG functional pathway for a given cluster gene set comprised of both mRNAs and lncRNAs.}
{\begin{tabular}{lrrrr}
\toprule
Genes & lncRNAs & KEGG Term & Overlap & P-value \\
\colrule
 ZNF177,ZNF175,ZNF607,ZNF606,... &                          AC022150.4 &      Herpes simplex virus 1... &  269/492 &     2.4e-323 \\
 OR7G2,OR8I2,OR7G1,OR9K2,OR11... &  AC131571.1,... &                Olfactory transduction &  350/444 &     2.4e-323 \\
 GSK3B,HDAC2,PTGER3,PTEN,... &                           LINC00598 &                    Pathways in cancer &   25/530 &      1.13e-13 \\
 CHRND,PTGIR,EDNRB,MTNR1B,... &                                UCA1 &  Neuroactive ligand-recept... &   12/338 &      9.67e-13 \\
      KIR2DS4,KIR2DL1,KIR3DL3,KIR2DL3 &  Z99756.1,... &   Antigen processing... &     4/77 &      2.62e-07 \\
 CHRNG,HTR1E,PTGER1,KISS1R,... &             AL355297.4,... &  Neuroactive ligand-recept... &    6/338 &      2.35e-06 \\
 P2RY4,TAAR6,C5AR1,HTR5A,TRHR &                          AC008125.1 &  Neuroactive ligand-recept... &    5/338 &      9.616-06 \\
                             AOC3,PAH &              LINC01940,... &              Phenylalanine metabolism &     2/17 &      2.19e-04 \\
 \botrule
\end{tabular}}\label{cluster_gsea}
\end{table*}

\begin{figure*}[t]
\begin{tabular}{c}
\centerline{\includegraphics[width=0.55\linewidth]{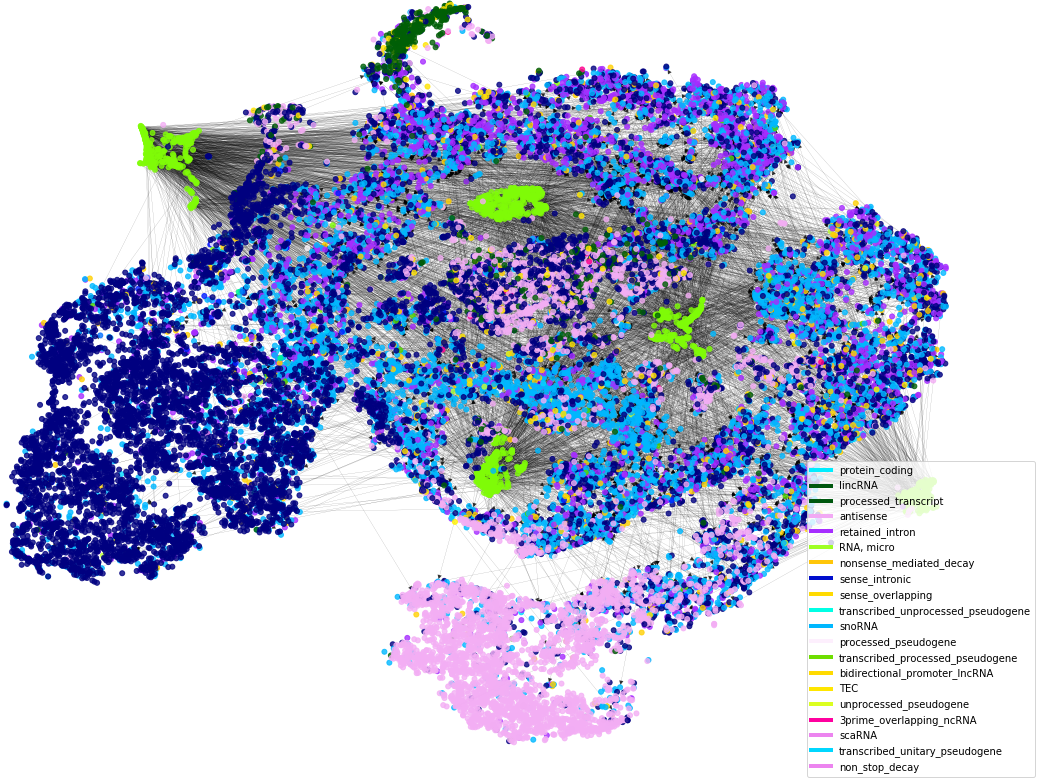}} \\
\end{tabular}
\caption{Visualization of the 35,069 transcriptome-wide lncRNA-miRNA-mRNA nodes mapped to a 2-D t-SNE projection from 128-dimensional embeddings. 
}
\label{lncRNA_subnetwork}
\label{tsne_viz_emb}
\end{figure*}

Additionally, since our method was able to map the functional affinity between RNA nodes belonging in disconnected components in the interaction topology, we hypothesize rna2rna can effectively map any RNA transcript sequence to a functional manifold in the embedding representation. It is observed in the t-SNE visualization in Fig. \ref{tsne_viz_emb} that the embeddings can preserve the local structure of the interactions and functional annotations, as well as exhibit good separation based on their transcript biotype classification. There is a clear separation between miRNAs to lncRNAs and mRNAs, and while the protein-coding mRNAs and lncRNAs may have some overlap, it is expected since the sequence structure of these two RNA classes is similar. Note that although no negative undirected edges between RNAs of different types (e.g., lncRNAs v. miRNAs) were sampled to explicitly designate different RNA types to have dissimilar embeddings, the network can still make a distinction between their functional roles. This shows that rna2rna's source-target embedding is an effective representation that can encode an RNA's biological function only by its given directed interactions.

Our main contribution proposes a highly versatile architecture aimed at predicting interactions between heterogeneous RNA transcripts while characterizing the functional landscape of various non-coding RNA species. The method currently has some limitations inherent in a deep learning method, such as requiring the tuning of free parameters and lacking direct biological interpretability of the embedding features. For future work, we can incorporate attention mechanisms in the encoder network to correlate individual features with specific regions in the sequence input. The notion of the source and target context embeddings would serve additional purposes to identify the binding regions of a sequence that would enable their regulatory functions.
In foresight, we intend this method to be the groundwork for subsequence down-stream analysis, where relevant attribute data and model parameters can be fine-tuned for specific prediction tasks such as the inference of a functional regulatory pathway or the discovery of unknown RNA families and novel disease biomarkers. Further development of this framework can provide an invaluable tool to support significant discoveries in systems biology, especially for newly identified non-coding RNAs.

\bibliographystyle{ws-procs11x85}
\bibliography{cites}

\end{document}